\documentclass[12pt, letterpaper]{article}

\usepackage{amsmath,amssymb}
\usepackage{cite}
\usepackage{fancyhdr}
\usepackage[top=1in, bottom=1in, left=1in, right=1in]{geometry}
\usepackage{graphicx}
\usepackage{hyperref}

\numberwithin{equation}{section}
\date{}

\begin{document}
\title{{\rm\footnotesize \qquad \qquad \qquad \qquad \qquad \ \qquad \qquad \qquad \ \ \ \ \ \                      RUNHETC-2024-31
}\vskip.5in    ``Observables" in de Sitter Quantum Gravity: in Perturbation Theory and Beyond}
\author{Tom Banks\\
NHETC and Department of Physics \\
Rutgers University, Piscataway, NJ 08854-8019\\
E-mail: \href{mailto:tibanks@ucsc.edu}{tibanks@ucsc.edu}
\\
\\
}

\maketitle
\thispagestyle{fancy} 

\begin{abstract}  A review of some errors made by the author and others in their search for quantum models of gravity in cosmological space-times that asymptote to de Sitter (dS) space in the future. The ``static de Sitter Hamiltonian", which measures the energy of localized objects in a static patch, is not a conserved quantity. The static time translation diffeomorphism of eternal dS space is a gauge symmetry, and the static energy is an approximate property of meta-stable constrained states.  It's not clear whether a theoretical model has to have a time independent Hamiltonian at all, but if it does, its eigenvalues are, {\it in principle}, not accessible to measurement by local detectors. \normalsize \noindent  \end{abstract}


\newpage
\tableofcontents
\vspace{1cm}

\vfill\eject
\section{Introduction}

Towards the end of the twentieth century it began to become clear that the universe we inhabit appeared to be evolving towards a dS space-time with a Hubble radius about $10^{61}$ in Planck units.  For those of us who were string theorists, this posed an acute problem.  Perturbative string theory seemed to make sense only in asymptotically flat space-time.  The discovery of the AdS/CFT correspondence allowed us to think about anti-dS space and certain mild perturbations of it (which did not perturb its asymptotic boundary), but there was nothing in string theory that looked anything like dS space.  We had of course known for a long time that string theory could not deal with the Big Bang or inflationary origin of our universe.  Many string theorists explained this to themselves by saying that "we were just doing particle physics, not cosmology", but the fact that all Poincare invariant string models had exact SUSY made that little myth a part of science fiction as well.

With W. Fischler, the present author first responded to this crisis by trying to imagine a model of eternal dS space.  We quickly realized that the thing that naively converged  to the Penrose diagram of Minkowski space was a single static patch of dS, and that most particles, thought of as geodesics, would asymptotically approach the horizon of the patch in a time of order the dS radius $R$.  We were of course familiar with the behavior of horizons, from extensive experience with black hole physics, and realized that there could be no asymptotic observables.   We therefore proposed that the observables in a dS universe would have to be finite time amplitudes, since the eventual fate of every state was to return to the equilibrium of the empty dS universe\cite{mtheoryobs}.

Fischler and Susskind\cite{fs} had earlier proposed an extension of the Bekenstein-Hawking\cite{BH} bound on black hole entropy to cosmological space-times, and Bousso\cite{raph} had generalized this to a bound on the entropy of a general causal diamond in a general space-time, in terms of the maximal $d-2$ volume ("area") of leaves of a null foliation of its boundary.  This led to the independent conjectures\cite{tbds}\cite{wfds} that the Gibbons-Hawking entropy of dS space was the logarithm of the dimension of the Hilbert space describing dS space in a quantum theory of gravity.  This implies that the density matrix of empty dS space is proportional to the unit matrix.  

This claim was met by incredulity, if not ridicule, by much of the community (including some people who now call it the "central dogma about dS space"), but there were a number of sanity checks on it.   The most important of these is that the introduction of localized objects into dS space {\it reduces the entropy}.  Indeed, for small mass of the localized object, the BGHJCSFSB\footnote{Bekenstein-Gibbons-Hawking-Jacobson-Carlip-Solodukhin-Fischler-Susskind-Bousso} entropy law gives a derivation of the Gibbons-Hawking temperature of dS space independent of quantum field theory. It invites an interpretation of that temperature as a statement that the localized energy is a measure of the number of constrained holographic q-bits on the boundary of the static patch.  It turns out\cite{tbpd} that this interpretation can be sustained even if the density matrix is {\it not} maximally uncertain.   An interpretation of localized states as living in a constrained subspace, gives a simple quantum mechanical explanation of why they should eventually return to the empty dS equilibrium state.  

Another vexing issue with the claim that the quantum dS space had a finite number of states was the non-compact nature of the dS isometry group.  In private conversations with L. Susskind, E. Witten, G. Horowitz and others, I often invoked the fact that most of the dS isometry group just mapped one static patch into gauge copies of itself.  Indeed, in the constructive field theory computation of non-gravitational global dS field theories from Euclidean path integrals\cite{nappietal} on the sphere, one uses precisely this mapping to prove dS invariance of the correlators.  In a gravitational theory, these mappings are gauge transformations.   What I missed in this argument was that dS transformations that {\it preserved} a given static patch are gauge transformations as well.

This fact was emphasized by Marolf and collaborators\cite{marolfetal}, based on earlier work\cite{earlier} and led to the procedure of "group averaging".   As emphasized in\cite{CLPW} it is most efficiently done by the standard procedure of BRST gauge fixing in a covariant gauge in perturbative quantum gravity.

The covariant BRST gauge fixing preserves the illusion of locality at the expense of introducing state spaces with indefinite metric.   As a consequence, the algebras of bounded functions of smeared local operators in the BRST quantized theory are not von Neumann algebras.  Nonetheless, it will still be true that commutators are singular on light cones, so that the full operator algebra does not factor into a tensor product of algebras localized in the two causally disconnected static patches.   As a consequence it will still be true that the static patch time translation operator acts as an outer automorphism on the local algebra in the static patch, and the naive "one sided static patch Hamiltonian" will have UV singularities.   The authors of\cite{CLPW} have argued that these can be removed by a version of the crossed product construction of von Neumann algebra theory, and that this is a step towards the correct quantum theory.  In a somewhat disconnected argument they have proposed that a projection on a Type $II_1$ sub-algebra of the standard Type $II_{\infty}$ crossed product is the correct algebra for dS space, so that they could adopt the proposal of\cite{tbds}\cite{wfds} that the density matrix was maximally uncertain.

In this paper, we will instead follow the standard BRST quantization procedure and ask what perturbative quantum gravity tells us about the "observables" of dS space.  Although our arguments will be somewhat heuristic, since we will be cavalier about the choice of gauge and the proper treatment of ghosts, we will argue that a picture similar 
to that of\cite{mtheoryobs} arises, except that the perturbative observables appear to be true asymptotic quantities localized at points on the boundary of a given static patch.  We then argue that the fast scrambling behavior of horizons\cite{lshpss} and the hypothesis that dS space has a finite dimensional Hilbert space imply that this localization is a temporary phenomenon and that the original picture proposed in\cite{mtheoryobs} is roughly correct.   In the final section we discuss the correct way to implement "modular flow" in the quantum theory of dS space, and in more realistic cosmological models, and propose an approximate meaning for "eigenvalues of the static patch Hamiltonian" within that definition of time evolution.

\section{Perturbative Quantum Gravity}

The BRST formalism for perturbative quantum gravity has been explained concisely in the second appendix of \cite{CLPW} .  One works in a gauge in which only the global dS isometries are unfixed (the precise gauge choice is not specified but has been discussed in earlier work by Marolf and collaborators and the papers they cite).  There are then sectors of ghost number ranging from $0$ to the dimension of the isometry group of dS.  The space of physical states is the BRST co-homology with maximal ghost number.   The algebra of BRST invariant operators is thus represented by functionals of the original fields that are invariant under the dS isometries.  Heuristically, these are integrals of local densities over dS space, divided by the volume of the dS group.  

As mentioned in the introduction, our treatment of the BRST formalism will be heuristic and based on the model of world sheet string theory, which similarly deals with a non-compact group of diffeomorphisms acting on a space-time with compact spatial sections. The role of the ghosts is primarily to properly define what it means to average over the non-compact group action\cite{marolfetal}.  The statement that a collection of operators all belongs to the algebra localized in a single static patch is invariant under diffeomorphisms in dS space, and consequently in perturbative quantum gravity expanded around the dS background.  dS group averaging of expectation values of products of such operators in the dS invariant vacuum state just corresponds to restricting attention to the operator algebra restricted to a given static patch.  

The most interesting question is how to average over the static time translation, which acts as modular flow in the local operator algebra.  Let's imagine an operator that transforms as a scalar density, localized around some particular point $(r,\Omega, t)$.   There is a unique radial geodesic through this point.  The diffeomorphism generated by static time translation moves the point along this geodesic.  For most of the infinite static proper time interval the transformed points are within a spacelike Planck distance from the horizon.  So it is plausible that a rigorous definition of the BRST invariant observable associated with this operator will be an operator $\phi_r (\Omega)$ localized around some point on the $d - 2$ sphere.  Of course, we also have to average over rotations, which are also gauge transformations.

However, if we now consider a product of different operators, located at separated points in the static patch, the rotational averaging will preserve the angular separation between those operators.

The conclusion of this non-rigorous analysis is that group averaging will lead to observable correlation functions of the form
\begin{equation} \langle \phi_{r1} (\Omega_1) \ldots \phi_{rN} (\Omega_N) \rangle , \end{equation} on the $d - 2$ sphere, which are invariant under global rotations.  Another way to think of these correlation functions is that they are functions on the space of time-like geodesics on dS space (whose boundary is the space of null geodesics) modulo the overall dS isometry group.  We can associate the observables to fields localized on gravitational Wilson lines along the geodesics. 

Of course, depending on the field content of the model, there may be many choices for the scalar densities $\phi$ that label these correlators.   In QFT, many of these densities might be related by operator product relations, but those relations are obscured by the fact that gravity is not renormalizable.   The list of "independent" scalar densities also depends on the "non-gravitational field content" of the effective field theory model.

String theory sheds light on both of these ambiguities.  In the Minkowski limit of dS space we expect the only perturbative observables correspond to scattering amplitudes of perturbative strings and D-particles (if any), whose spectrum is determined by the choice of compact background.  Away from the weakly coupled string regime, but still assuming gravity can be treated perturbatively, we would expect only a small number of distinct amplitudes, corresponding to particles carrying conserved quantum numbers, would survive.   

Each of these operators is the sum of two terms, one from the past and one from the future boundary of the static patch.  The discrete diffeomorphism of static time reflection allows us to separate out even and odd (and thus past and future) pieces of these individual operators, so each of these correlators can be broken up into sums of terms, one of which will be a pure transition operator between past and future.  In\cite{harlow} it was argued that one must gauge all reflections that are true symmetries of dS space.   Static time reflection, combined with some operation on the fields in the effective field theory to make TCP, will always be a symmetry and should be gauged.  However, we can still use {\it relative} time reflections between different geodesics to separate out transition amplitudes from correlations between fields purely on the future or past horizons.

These perturbative observables resemble the finite time transition amplitudes proposed in\cite{mtheoryobs}.  The difference is that the very sketchy discussion in that paper was groping towards the idea that was most fully expressed in\cite{lshpss}: finite area horizons are fast scramblers of quantum information and the system returns to equilibrium in a time of order $R {\rm ln}\ (R/L_P)$.  Note that the factor of $R$ in this estimate is observer dependent and refers to proper time as measured on a trajectory that is far from the horizon.  The logarithm is the log of the entropy, which is a power of $R/L_P$, and there is a system dependent numerical coefficient in the estimate which is not easy to calculate for a given fast scrambling Hamiltonian.  This implies that there can be no true asymptotic observables\cite{sussetal}, which was the intuition expressed in\cite{mtheoryobs}.   The formalism of strictly perturbative quantum gravity cannot detect these non-perturbative features, which are nonetheless clearly visible in classical gravitational physics.  The simplest philosophical explanation of that fact is Jacobson's\cite{ted95} observation that Einstein's field equations are the hydrodynamic equations of the area law for entropy.  Hydrodynamics can sometimes be used as a classical field theory, to be perturbatively quantized in order to reveal low energy excitations of a gapless ground state, but it also describes the behavior of high entropy states, whose microscopic dynamics is too complex to follow in detail.  Hydrodynamics does give a coarse grained description of departures from equilibrium and the return to the equilibrium state.  In the dynamics of horizons this is the role of the familiar quasi-normal modes.

\section{The Meaning of the Static Hamiltonian}

The present author, as well as many other researchers who've tried to come up with a quantum theory of dS space, has spent a lot of time and print on the properties of "the Hamiltonian that evolves the system in static patch time".  Yet the idea of localized objects as constrained states of a system that constantly tends to equilibrium on time scales of order $R {\rm ln}\ (R/L_P)$ denies the possibility of an actual conservation law associated with this operator.   Indeed, classical geodesics, with a single exception, approach within a space-like Planck separation of the horizon of the patch, in a time of order $R$.  Quantum wave functions of single particle relativistic wave equations, starting from finite time initial conditions concentrated around the static geodesic, spread uniformly over the horizon in times of order $R$, up to logarithms.   

In a previous paper\cite{hilbertbundles}, I proposed a general prescription for talking about time evolution in finite causal diamonds.  It is based on a Jacobsonian interpretation of a classical space-time background as a hydrodynamic description of a quantum system, and is {\it by definition not background independent}.  The arena for quantum gravity is a bundle of Hilbert spaces over the space of time-like geodesics of the hydrodynamic background.
In each Hilbert space one has a separate quantum system, which evolves in the proper time along that geodesic.  Nested intervals of proper time define nested causal diamonds in the space-time and we assign a tensor factor of fixed dimension in the Hilbert space of a fiber, to each diamond according to what will be called the Carlip-Solodukhin (CS) rule\cite{carlip}\cite{solodukhin}\footnote{In order for the notion of tensor factor to be well defined for non-positive c.c., we must restrict the proper time intervals to diamonds of finite area. Afterwards we can take the infinite area limit with care.  In the AdS case this would mean renormalization of a holographic tensor network description to give a continuum field theory on the boundary. For zero c.c. it means determining the non-perturbative Hilbert space on which the scattering operator acts, a problem that I consider incompletely resolved even in high dimension.}.   Strict locality is imposed by insisting that the evolution operator be time dependent in a way that preserves this tensor factorization at every time step.  The CS ansatz implies that each tensor factor Hilbert space is finite dimensional, so we have an analog of Algebraic Quantum Field Theory (AQFT) with Type $I_N$ algebras instead of Type $III_1$.   It's well known that sharp locality and time independent Hamiltonians imply Type $III_1$, so the CS rule imposes both evolution by discrete time steps and that the Hamiltonian {\it must} be time dependent.   A geometric way of seeing this is that coordinate systems that remain inside a causal diamond\cite{CHM} are never generated by time-like Killing vectors, even when the global space-time has such isometries.  

In QFT, the modular operator $\Delta_{\diamond} = e^{ - K} $ of a causal diamond is a positive operator in the Hilbert space which is not a member of the algebra of operators localized in the diamond.  The modular flow 
\begin{equation} a \rightarrow e^{i K t} a e^{-i K t} , \end{equation} for $a \in {\cal A}_{\diamond}$ is an outer automorphism of the local algebra.  One reason the modular operator is interesting is that the operator
\begin{equation} U(t,\delta) = e^{i K(t)} e^{- i K(t + \delta)} , \end{equation} generates an evolution on the operator algebra of a diamond corresponding to a proper time interval $[-T,t + \delta]$ along some geodesic, which, when $\delta \ll t + T$, maps the subalgebra localized in the subregion of that diamond not contained in the diamond corresponding to the interval $[-T, t]$, into itself.  That is, we can think of $U(t,\delta)$ as an infinitestimal time evolution operator on time slices interpolating between nested diamond boundaries.

According to the CS ansatz, local algebras are finite dimensional, so $K$ is an operator in the local algebra.  CS (as generalized by\cite{BZ}) argue that it is the $L_0$ generator of a cutoff $1 + 1$ dimensional CFT, with central charge proportional to the diamond area, so that Cardy's formula matches the area law.  This ansatz also predicts a fluctuation formula
\begin{equation} \langle (K - \langle K \rangle)^2 \rangle = \langle K \rangle , \end{equation}
that was shown to be valid for Ryu-Takayanagi diamonds in {\it every} AdS/CFT model in the Einstein-Hilbert limit\cite{VZ2}\cite{perl}\cite{deBoer}.  Our remark about modular operators in QFT motivates the definition of discrete time evolution between nested causal diamonds with Planck time separated future tips as\footnote{We use notation for the initial point in proper time appropriate to a cosmological space-time.  The modifications necessary for a time symmetric situation are straightforward.}
\begin{equation} U_{in} (t,0) = e^{i L_0 (t)} e^{-i L_0 (t + L+P)} . \end{equation}   This is an operator that unitarily embeds the Hilbert space of the smaller diamond into the larger one.  It must be supplemented by an operator $U_{out} (t,0)$ which operates on the tensor complement of the diamond Hilbert space in the full Hilbert space of the geodesic.

The fundamental unproven conjecture of this approach to quantum gravity is that $U_{out} (t,0)$ can be consistently determined for all trajectories and all times by imposing {\it The Quantum Principle of Relativity} (QPR), which plays the role of a connection on this bundle of Hilbert spaces.   Each pair of diamonds along each pair of trajectories, has an overlap, which contains a maximal area causal diamond.  The QPR states that the dynamics and initial conditions on all fibers of the Hilbert bundle must be chosen in such a way that the density matrix on any overlap, as computed along one trajectory, has the same entanglement spectrum as that computed along the second trajectory.  These conditions are easy to state, but it has proven hard to develop a formalism that makes them easy to satisfy.  It's unclear whether this is a consequence of the incompetence of the small number of developers of the formalism or the intrinsic difficulty of the problem.

Within this formalism, we can now begin to understand the meaning of the static Hamiltonian in the causal patch of dS space.   The Hilbert space along any geodesic is finite dimensional and the density matrix representing the empty causal patch is the CS density matrix with central charge determined by the Gibbons-Hawking entropy.  States with total energy $E$ localized "near" the geodesic lie in a subspace whose projector has expectation value $e^{- 2\pi ER}$ in the CS density matrix.  As emphasized in\cite{tbpd}, this is valid for all states with $ER \gg 1$, but only for an order one entropy of states with $ER \sim 1$.  Most of the states with $ER \sim 1$ can neither be described in terms of QFT, nor as meta-stable black holes at rest at the origin.  The description of dS space in terms of fermion matrix models\cite{bfm} (updated to fermionic matrix CS-CFTs in\cite{hilbertbundles}) views these constrained subspaces (for space-time dimension $d = 4$) as subspaces in which $E\times R$ blocks of fermion bilinears are set to zero, leading to decoupling of the diagonal blocks in a single trace Hamiltonian.   The static geodesic proper time scale for equilibration of those frozen variables is $C R\ {\rm ln}\ (ER)$\footnote{The constant $C$ depends on the details of the fast scrambling Hamiltonian, and is not easy to calculate.}, so we can understand why $E$ appears to be conserved in the classical limit $(R/L_P) \rightarrow \infty$, and also why it becomes a true asymptotic symmetry in the $R\rightarrow\infty$ limit of Minkowski space.  $E$ defined in this way is not a conserved quantum number in dS space.   

When $R/L_P$ is large, and $E$ describes long lived excitations, we should also be able to have an approximate notion of localization of energy in the static $r$ coordinate, corresponding to the fact that the perturbative observables carry the label $r$ denoting the closest approach to the origin of the geodesic that defined the gauge invariant observable.  

The formalism of\cite{hilbertbundles} provides two complementary ways to talk about localization of energy.  It describes the static diamond in eternal dS space as a time symmetric nested sequence of diamonds, as shown in Figure 1.  
\begin{figure}[h]
\begin{center}
\includegraphics[width=01.5\linewidth]{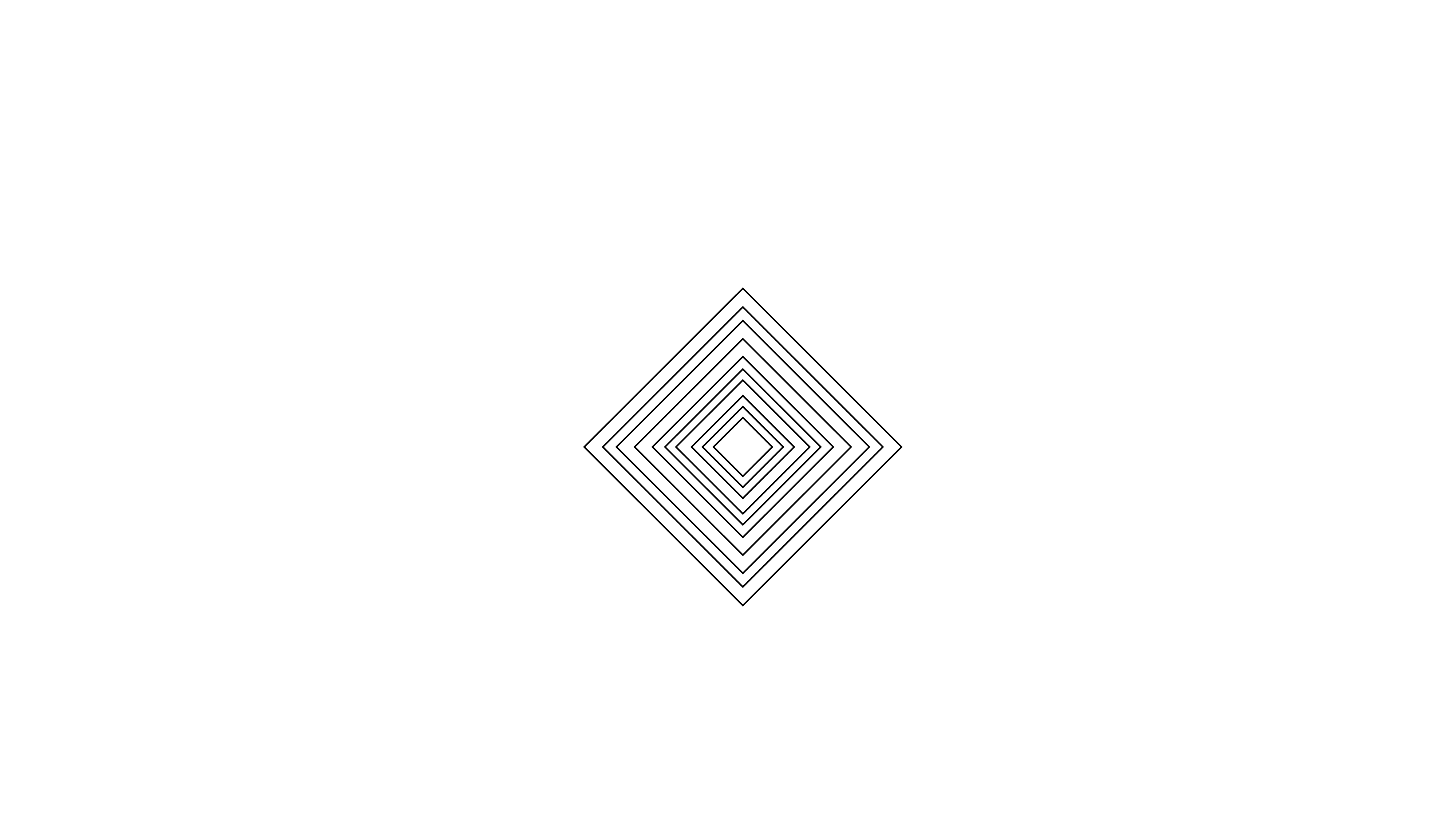}

\caption{A Time Symmetric Nested Cover of the Causal Patch by Causal Diamonds. The Time Separations Are Planck Scale} 
\label{fig:symmetricnestedcoverofadiamond}
\end{center}
\end{figure}

\begin{figure}[htt]
\begin{center}
\includegraphics[width=0.5\linewidth]{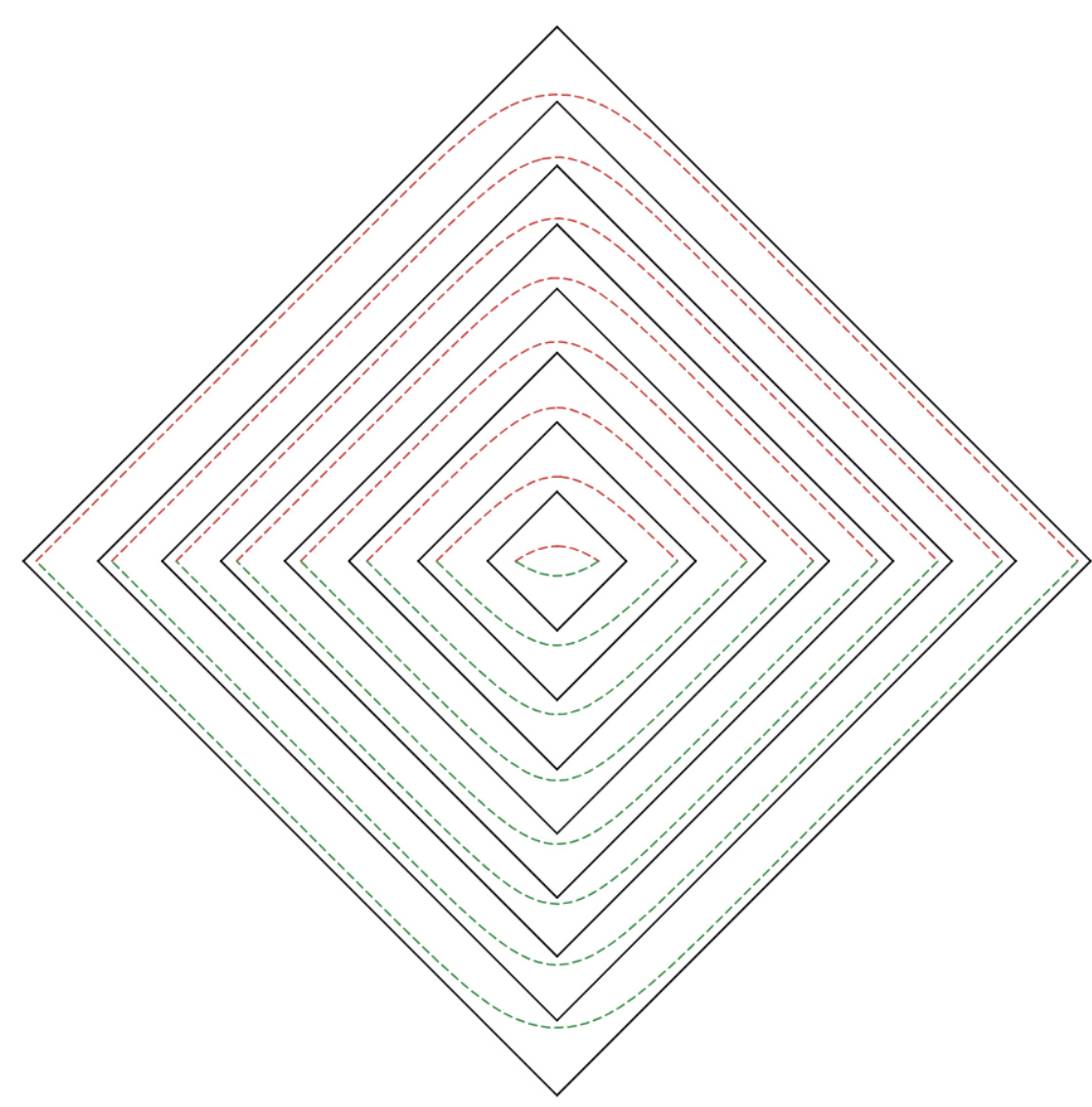}

\caption{A Time Symmetric Nested Cover of the Causal Patch by Causal Diamonds With Natural Timeslicing for Redshift Distance Relations} 
\label{fig:symmetricnestedcoverofadiamondwchmlines}
\end{center}
\end{figure}

Localized excitations are sequences of constrained states on the variables in each of those diamonds and the existence of those constraints defines a correlated set of angular coordinates on the nested holoscreens.  These should be identified with gravitational Wilson lines in the effective field theory, following the geodesics that described the gauge invariant field theory observables\footnote{Those readers worried about the usual quantum fluctuations of "particle trajectories" in field theory, should recall that above three dimensions particles are inevitably accompanied by jets of soft gravitons.  The constraints of\cite{bfm}\cite{tbwfsmatrix}\cite{hilbertbundles} can be thought of as "jet isolation criteria" that separate the soft gravitons accompanying hard momentum fluxes from the more uniformly distributed\cite{weinberg}\cite{venezianoetal} states that populate most of the boundary of all causal diamonds. The microscopic description of these states by bulk QFT is incorrect if one believes the C-S ansatz, although QFT is perfectly adequate for calculating inclusive cross sections, which ignore everything but the total energy pumped into these states in Minkowski space.}  .    So the localization of states in the $r$ coordinate should be enforced by imposing the QPR on time evolution.   Constraints must be imposed on the "out" Hilbert space of the description of the system along one geodesic fiber of the Hilbert bundle, to match those imposed on the "in" Hilbert space of a different geodesic whose closest approach to the original of the first static patch is at $r$.

In order to describe these constrained states consistently with the QPR, the Hamiltonian $H_{in} (t) + H_{out} (t)$ in a given fiber, which describes propagation along the geodesic at the origin of the static patch, must involve a sum of terms, {\it which depend on the initial conditions on the past boundary of the patch}.  These determine, in terms of the "gravitational Wilson lines" , how many such terms there are, and {\it their relative normalizations}.   The ansatz that has been proposed is that the extra terms just sit in the appropriate diagonal blocks of the matrix Hamiltonian and consist of a constant denoting the energy of the excitation and a copy of the full matrix Hamiltonian, scaled down to the smaller block.  The overall normalization of these terms is determined by the "redshift-distance relation" in the coordinate system sketched in Figure 2.   

The non-constant term in the Hamiltonian of a block is responsible for the emission and absorption of soft particles from the excitation.  These correspond to processes in which the block spontaneously breaks up into a smaller block and a larger one and reverts to generic size.

For each value of proper time along the geodesic, the terms in the Hamiltonian are distributed between $H_{in} (t)$ and $H_{out} (t)$ according to the configuration of the gravitational Wilson lines.   By analogy with Jacobson's identification of the background solution of Einstein's equations as the hydrodynamic flow of an underlying quantum system, we can view the gravitational Wilson lines as the hydrodynamic trace of the propagation of localized excitations through the bulk.  The fundamental microscopic definition of this process is the evolution of certain constrained states and we must construct the Hamiltonian of the constrained subspace in order to match the hydrodynamic information.  An interesting feature of this ansatz is that we see the necessity for choosing a set of gravitational Wilson lines to describe a scattering event whose microscopic description will involve jets of soft particles collimated along those lines, surrounded by isolation zones separating the jets from a holographic ensemble of near horizon states with "energies" of order $1/R$.

In\cite{hilbertbundles} we provided a complete ansatz for $H_{in} (t)$ and conjectured that $H_{out} (t)$ would then be determined by the QPR.  The QPR is particularly powerful for eternal dS space, because $H_{in} (t)$ has to be identical for every geodesic in the system, and only the boundary conditions defined by the Wilson lines distinguish one geodesic from another.  The unsolved problem in this approach to quantum gravity is a definition of $H_{out} (t)$ that will solve the conditions imposed by the QPR.  The lesson from string theory is that this is likely to be difficult.  Most low energy effective field theories of gravitation do not come from consistent string models with either asymptotically flat or AdS boundary conditions.   We have no reason to expect that consistent models of dS space are any easier to find, especially since the $R/L_P \rightarrow \infty$ limit of such a model is likely to lead to a consistent model of quantum gravity in Minkowski space.   

In\cite{tbwfsmatrix} we showed that the matrix form of $H_{in} (t)$ naturally led to the right parametric form of the Newtonian gravitational law for the leading long distance interaction between a pair of localized excitations in a causal diamond of size $r$.   The interaction is generated causally, by excitation and de-excitation of the frozen q-bits connecting the small blocks of the matrix to the large block of excitations on the diamond boundary.  We also argued that the QPR implied that there would be "time ordered Feynman diagram" like exchange interactions.   In the case of graviton graviton scattering, in a Lorentz covariant theory, we know that these two contributions exactly cancel when the graviton momenta are parallel.   One can see that this will not happen in general in these matrix models, so satisfying the QPR constraints is indeed a non-trivial 
unsolved problem.   

\section{Conclusions}

We have argued, following\cite{marolfetal} that a quantum theory of dS space should not be invariant under any of the classical isometries of the space-time, which are instead merely gauge transformations.  We've proposed a heuristic identification of perturbative gauge invariant observables in terms of correlation functions of operators on the boundary of a single dS static patch, invariant under rotations.  The operators are labeled by radial geodesics, local scalar densities in the effective field theory, and a discrete quantum number distinguishing past from future static time.  A more rigorous investigation of the BRST cohomology of perturbative quantum gravity in dS space should be done to verify this conjecture.

We then argued that the hypothesis\cite{tbds}\cite{wfds} that the Hilbert space of the quantum theory is finite dimensional, particularly when combined with the {\it fast scrambling} conjecture\cite{lshpss}, implies that none of these perturbative observables survives in the actual quantum theory.  Rather, as first suggested in\cite{mtheoryobs} they correspond to finite time transition amplitudes for constrained out of equilibrium states, which return to the empty dS equilibrium in a time of order $R {\rm ln}\ (R/L_P)$.   

Finally we discussed the holographic space-time formalism\cite{hst}, which has taken on a new form in\cite{hilbertbundles}, and in which the "static Hamiltonian" could be given an approximate meaning as an operator in certain constrained meta-stable states of the system.   As it stands, the HST formalism describes the final equilibrium state of dS space as evolving under a time independent Hamiltonian.  The eigenvalues of this Hamiltonian do not correspond to any energy measured by local detectors.  Local energies are related to the dimensions of constrained subspaces to which localized states belong, and the times that it takes them to return to equilibrium\footnote{These are proper times as measured by a detector at the origin.}.  The energies of near horizon states are all of order $R^{-1}$ and typical level spacings are of order $e^{- \pi (R/L_P)^2} R^{-1}$ .  One suspects that there is a lot of freedom to add random time dependence into this model without appreciable effect on the physics of localized excitations.   \cite{hilbertbundles} proposed a particular Thirring interaction between a set of cutoff two dimensional fermion fields $\psi_a$ as a candidate for the CS $L_0$ generator.   The label $a$ runs over the spinor spherical harmonics $Y_a (\Omega)$ on the two sphere, with an angular momentum cutoff determined by the CS relation between the central charge of the near horizon CFT and $\pi (R/L_P)^2$.   Define the field
\begin{equation} \psi (z,\Omega) = \sum^{a_{max}} \psi_a (z) Y_a (\Omega) .  \end{equation}   Then the local 2d interaction density of the Thirring model is 
\begin{equation} {\cal L}_{int} (z) = g \int d^2 \Omega [ \bar{\psi} (z,\Omega) \gamma^m \psi (z,\Omega) \bar{\psi} (z,\Omega)\gamma_m \sigma_3 \psi (z,\Omega)] . \end{equation} 
When discussing time evolution in diamonds much smaller than the dS radius, the Hamiltonian is time dependent because we're constantly adding degrees of freedom to the "in" part of the system.  Since the evolution is discrete, this must stop at some finite time when the Hilbert space reaches its maximum dimension, but we can still allow the Thirring coupling $g$ to vary with time.  Once we've given up the mistaken idea that we must impose the global isometry of the static patch as a global symmetry of the system, there is no reason not to do so.  It seems clear that there are a large number of choices of time evolution which will not disturb the conclusion that detectors near the origin will experience "thermal physics at the fixed dS temperature", despite the time dependence of the underlying holographic variables.   From a philosophical point of view, this kind of time dependent Hamiltonian solves the silly problem of "Boltzmann Brains".  

We conclude by discussing an aspect of the universe we inhabit, which knowledgeable readers may find disturbing in light of the preceding discussion.  That universe appears to be approaching an asymptotically dS future with $ R \sim 10^{61} L_P$.   If it indeed has a positive cosmological constant of the indicated magnitude, then in about $100$ times its current age of $\sim 13$ billion years, all we will see in the sky is the local group of galaxies.   The time scale for the disappearance of everything else into the cosmological horizon is consistent with the fast scrambling estimate, but the local group will persist until it collapses into a black hole and radiates its quantum information back to the horizon. The time scale for this is {\it much} longer than the scrambling time.  

The answer to this conundrum is that the local group of galaxies is not traveling on a static geodesic\cite{dS3}.  Its trajectory is complex, perturbed by its interaction with everything else in the universe and it is constantly emitting radiation, both electromagnetic and gravitational, so neither its energy or angular momentum are conserved on any long time scale.  The trajectory of its center of mass is a collective coordinate of the huge quantum system made up of its constituents.  It is a completely decoherent classical variable, and the local group has many collective variables with similar properties.  The persistence of semi-classical physics in dS space, for detectors in the local group, depends only on the continued existence of these variables.

In summary, if a more careful BRST analysis confirms our heuristic picture of perturbative quantum gravity observables in dS space, then the additional assumptions of a finite dimensional Hilbert space and fast scrambling imply that there are NO asymptotic observables, as first suggested in\cite{mtheoryobs}\cite{sussetal}.  Instead, in a model of eternal dS space, there would be finite time transition amplitudes, for time scales at most of order $R\ {\rm ln}\ (R/L_P)$, very similar to the actual amplitudes we measure in real world laboratories as stand ins for the S matrix elements of quantum field theory.

For time scales longer than the scrambling time, the existence of detectors and sensible measurements of quantum systems in dS space depends on the existence of large complex bound states like star clusters and the details of what can be measured will, to some extent, depend on the individual structure in which the detector is embedded.  There does not seem to be much use for a model of the detector as an isolated elementary system traveling on a geodesic in a static patch.  If it's truly elementary, its wave function will spread over the horizon in less than a scrambling time.  If it's a black hole, it has no useful pointer variables with which to make complex measurements. Only a complicated, uncollapsed gravitational bound state has the resources to host a measuring device.  Such a system will not travel on a geodesic, and will constantly emit radiation.  

In\cite{hilbertbundles} we presented a formulation of quantum gravity in which the entire dS group acts on a bundle of Hilbert spaces, with the stabilizer subgroup of a geodesic acting on the Hilbert space of a particular fiber.  The question we have addressed in this paper is whether that second action has anything to do with measurements of observable energies and angular correlations in real world physics.  The answer as far as energy is concerned is definitely no.  The energies that we measure in scattering experiments near a static geodesic are related to the number of constraints on the holographic variables, rather than to eigenvalues of their asymptotic Hamiltonian.  We've even speculated that it would be possible to give the Hamiltonian in the asymptotic finite dimensional Hilbert space a sufficiently mild random time dependence, without affecting any local measurement.  

On the other hand the finite time transition amplitudes which we believe are the only true local observables in dS space do depend on relative angles, and if we lived in an eternal dS space they would be exactly rotationally invariant functions of relative angles, as a consequence of the underlying isometry.  In the real Big Bang universe we inhabit, this symmetry is preserved and evident in the statistics of the microwave sky.  Those statistical fluctuations also exhibit an approximate dS symmetry that is related to the properties of the very early universe, and logically distinct from the isometry of our apparent asymptotic future.

\vskip.3in
\begin{center}
{\bf Acknowledgments }
\end{center}
 The work of T.B. was supported by the Department of Energy under grant DE-SC0010008. Rutgers Project 833012.  Conversations with Patrick Draper about the material in this note are gratefully acknowledged.  Sidan A and Willy Fischler helped with the figures.  Fischler has of course been a long time collaborator on much of the material reviewed in this note but is not responsible for any errors I may have added.




\end{document}